\documentstyle[version2,aps]{revtex}\let\Style=3
\def\Bbb#1{{\cal #1}}
\makeatletter
\def\footnote{\@ifnextchar[{\@xfootnote}{\stepcounter
   {\@mpfn}\xdef\@thefnmark{\thempfn}\@footnotemark\@footnotetext}}
\def\footnotesize{\@setsize\footnotesize{9.5pt}\ixpt\@ixpt
\abovedisplayskip 6\p@ plus2\p@ minus4\p@
\belowdisplayskip \abovedisplayskip
\abovedisplayshortskip \z@ plus\p@
\belowdisplayshortskip 3\p@ plus\p@ minus2\p@
\def\@listi{\leftmargin\leftmargini
\topsep 3\p@ plus\p@ minus\p@\parsep 2\p@ plus\p@ minus\p@
\itemsep \parsep}}
\makeatother

\def\references{%
  \bgroup\narrowtext\linewidth=\columnwidth\vskip24pt%%
    \hrule width\columnwidth%
  \vskip1.6cm%
  \ifdim\baselinestretch in>1 in\parsep4pt\else\parsep 0pt\fi%
  \ifdim\baselinestretch in>1 in\itemsep4pt\else\itemsep 0pt\fi%
  \list%
  {{[\arabic{enumi}]\hskip4pt}}%
  {\settowidth%
     \labelwidth{\hbox spread4pt{\hskip\WidestRefLabelThusFar\hfil}}%
    \leftmargin\labelwidth%
    \labelsep=0pt%
    \advance\leftmargin\labelsep%
    \usecounter{enumi}}%
  \def\newblock{\hskip .11em plus .33em minus -.07em}%
  \sloppy%
  \sfcode`\.=1000\relax\small%
}
\def\endreferences{%
  \endlist\vskip1sp\egroup
  \gdef\@SetMaxRefLabel##1{}%
}

\def\narrowtext{\par\global\columnwidth20.5pc
  \global\hsize\columnwidth\global\linewidth\columnwidth
  \global\displaywidth\columnwidth
}
\def\widetext{\par\global\columnwidth42.5pc
  \global\hsize\columnwidth\global\linewidth\columnwidth
  \global\displaywidth\columnwidth
}

\newlength{\Kluge}\newlength{\MarkKluge}\newlength{\Mark}
\def\MarkColumn{
  \null\vspace{\Kluge}
  \newbox\partialpage
  {\output={\global\setbox\partialpage=\vbox{\unvbox255}}\eject}
  \Mark=\ht\partialpage
  \addtolength\Mark{\MarkKluge}
  \vbox{\unvbox\partialpage}
}
\def\Eject{\eject\null\vspace{\Mark}}

\expandafter\ifx\csname Style\endcsname\relax % (no \Style, i.e. REVTeX 2.0)
\else
  \if\Style0 % (preprint)
    \def\widetext{}\def\narrowtext{}\def\twocolumn{}
    \def\MarkColumn{}\def\Eject{}
  \else % (\Style must be 3, i.e. REVTeX 3.0)
  \fi
\fi

\def\bar{\overline}
\def\XX{{\scriptscriptstyle X}}
\def\AtSigma{|_{\scriptscriptstyle\Sigma}}
\def\OnSigma{_{\scriptscriptstyle\Sigma}}
\def\Uin{u^{\,\raise2pt\hbox{$\scriptstyle\rm in$}}}
\def\Uout{u^{\,\raise2pt\hbox{$\scriptstyle\rm out$}}}
\def\sgn{{\rm sgn}}
\def\p{\partial}
\def\mpm{{M^{\pm}}}
\def\upm{{U^{\pm}}}
\def\upms{{\scriptscriptstyle\upm}}
\def\ypm{{\mbox{\boldmath $\Theta$}^{\pm}}}
\def\smh{\sqrt{\vert h\vert}}
\def\d{{\mbox{\boldmath $\delta$}}}
\def\fp{F^+\AtSigma}
\def\fm{F^-\AtSigma}
\def\s{\Sigma}
\def\bff{{\mbox{\boldmath $F$}}}
\def\bfy{{\mbox{\boldmath $\Theta$}}}
\def\pp{\Phi^+}
\def\pmm{\Phi^-}
\def\ppm{\Phi^\pm}
\def\up{{U^+}}
\def\um{{U^-}}
\def\ll{{\delta}}

\twocolumn
\begin{document}
%draft

%**end of header

\widetext

%\rightline{\today}
\rightline{1 March 1993}
\rightline{gr-qc/9303002}
\null\vspace{-10mm}

\begin{title}
\centerline{\bf THE SCALAR FIELD EQUATION}
\centerline{\bf IN THE PRESENCE OF SIGNATURE CHANGE}
\end{title}
\null\vspace{-7mm}

\author{Tevian Dray}
\begin{instit}
\centerline{Department of Mathematics, Oregon State University,
                Corvallis, OR  97331, USA}
\centerline{\tt tevian{\rm @}math.orst.edu}
\end{instit}
\null\vspace{-10mm}

\moreauthors{Corinne A. Manogue}
\begin{instit}
\centerline{Department of Physics, Oregon State University,
                Corvallis, OR  97331, USA}
\centerline{\tt corinne{\rm @}physics.orst.edu}
\end{instit}
\null\vspace{-10mm}

\moreauthors{Robin W. Tucker}
\begin{instit}
\centerline{Department of Physics, University of Lancaster,
                Bailrigg, Lancs. LA1 4YB, UK}
\centerline{\tt rwt{\rm @}v1.ph.lancs.ac.uk}
\end{instit}

\null\vspace{-1.75mm}

\begin{abstract}
We consider the (massless) scalar field on a 2-dimensional manifold with
metric that changes signature from Lorentzian to Euclidean.  Requiring a
conserved momentum in the spatially homogeneous case leads to a particular
choice of propagation rule.  The resulting mix of positive and negative
frequencies depends only on the total (conformal) size of the spacelike
regions and not on the detailed form of the metric.  Reformulating the problem
using junction conditions, we then show that the solutions obtained above are
the unique ones which satisfy the natural distributional wave equation
everywhere.  We also give a variational approach, obtaining the same results
from a natural Lagrangian.
\end{abstract}

%\pacs{04.20.Cv,02.40.+m}

\narrowtext

\MarkColumn

\section{INTRODUCTION}

In previous work
\cite{PaperI}
we argued that signature change of a spacetime metric should lead to particle
production by determining the junction conditions on the scalar field.
A detailed  consideration of quite general propagation rules was given in
\cite{PaperII}, where the presence of symmetry was invoked to demand a
conserved momentum, thus singling out the propagation rule proposed in
\cite{PaperI}.  In this paper we give a mathematically cleaner presentation
of the result that a conserved momentum leads to a particular junction
condition on the scalar field.  We also propose a generalization using
distributional language which could be applied in a more general setting.

In Section II we establish our notation and then introduce our homogeneous
signature-changing model in Section III.  In Section IV we show that the added
physical requirement that momentum be conserved determines, using Stokes'
Theorem, the propagation of the scalar field across a surface of signature
change.  In Section V we reformulate the theory in terms of distributions,
deriving the natural distributional wave equation without invoking any
symmetry, and show that solutions of this wave equation automatically satisfy
the propagation condition above.  In Section VI we again reformulate the
theory, this time using a variational approach, and show that a natural choice
of action is equivalent to the distributional wave equation of the previous
section.  The results in Sections V and VI do not require the assumption used
in Section IV that the momentum be conserved, but instead derive momentum
conservation as a consequence of the theory.  Finally, in Section VII we
discuss
our results, contrasting the various formulations.

\Eject
\section{NOTATION}

We first review the usual theory of the massless scalar field equation in the
absence of signature change using the language of differential forms.
We set up our formalism on an $n$-dimensional manifold and then apply it to a
particular 2-dimensional model.
Associated with any closed $(n-1)$-form $\alpha$ there is an
integral conserved quantity obtained from Stokes' Theorem, namely
\begin{equation}
0 = \int_V d\alpha = \int_{\partial V} \alpha
\label{Stokes}
\end{equation}
It is therefore useful to express the theory in terms of forms.

The Lagrangian $\cal L$ for the massless scalar field with respect to an
arbitrary metric $g_{ab}$ is given by%
\footnote{For simplicity we assume in this section that $\Phi$ is real.}
\begin{equation}
{\cal L}\,{*}1 = d\Phi \wedge *d\Phi
\end{equation}
where $*$ denotes the Hodge dual, from which one derives the wave equation
\begin{equation}
d{*}d\Phi=0
\label{Wave}
\end{equation}
or, in tensor language, $\Box\Phi=0$.  By virtue of (\ref{Wave}), the
$(n-1)$-form
\begin{equation}
K = \Phi\, {*}d\Psi - \Psi\, {*}d\Phi
\end{equation}
is closed ($dK=0$) for any 2 solutions $\Phi$ and $\Psi$ of (\ref{Wave}).
In tensor language, the associated conserved quantity is just the symplectic
product (from which the Klein-Gordon product is constructed), namely
\begin{equation}
0 = \int_{\partial V}K
    = \int_{\partial V} n^a
        \left( \Phi \partial_a\Psi - \Psi \partial_a\Phi \right) \, d\Sigma
\end{equation}
where $n^a$ is the unit normal and $d\Sigma$ the volume element on
$\partial V$.
Finally, associated with any Killing vector $X$ there is a conserved current
given by the closed form
\begin{equation}
J_\XX = i_\XX \,d\Phi \wedge *d\Phi + d\Phi \wedge i_\XX \,{*}d\Phi
\end{equation}
where $i_\XX$ denotes interior product, so that for example
$i_\XX(d\Phi)=d\Phi(X)=X(\Phi)$. In tensor language, $J^a = T^{ab} X_b$ is
conserved due to the conservation of the stress-energy tensor $\nabla_m T^{ma}
= 0$ and Killing's equation $\nabla_{(a}X_{b)}=0$.

\section{SIGNATURE CHANGE}

Consider the manifold $M={\Bbb R}\times{\Bbb S}$ with metric
\begin{eqnarray}
\nonumber
ds^2 &=& f(t) \, dt^2 + g(t) \, dx^2 \\
                &=& g \left( h(t) \, dt^2 + dx^2 \right)
\label{Metric}
\end{eqnarray}
where $x$ is periodic, $h=f/g$, $g$ is everywhere positive, and we assume that
$f$ (and hence $h$) has at least one and at most countably many isolated roots
$\{t_0,t_1,...\}$.  For the remainder of this section, we will assume that $f$
has only one root, which occurs at $t=0$, and that $\sgn(f)={\rm
sgn}(t)$.  Note that the vector $X=\partial_x$ is a Killing vector.

Introduce new ``time'' parameters $\tau$ for $t<0$ and $\sigma$ for $t>0$
by
\begin{equation}
\tau = \int_0^t \sqrt{-h} \,dt \qquad\qquad
  \sigma = \int_0^t \sqrt{h} \,dt
\end{equation}
so that, away from $t=0$, the metric takes the form
\begin{equation}
ds^2 = g \left( \sgn(h) \,dT^2 + dx^2 \right)
\end{equation}
where $T$ is $\tau$ or $\sigma$ as appropriate.
Note that while the conformal ``time'' parameter $T$ is continuous, it is not
$C^1$ related to $t$ and thus cannot be used as a coordinate in a region
that includes $t=0$.

Away from $t=0$, it is easy to find a (complex) basis of solutions of
(\ref{Wave}) using conformal coordinates, namely
\begin{eqnarray}
\nonumber
u_k = & e^{ikx}e^{-i|k|\tau} \qquad & (h<0)\\
           v_k = & e^{ikx}e^{-k\sigma} \qquad & (h>0)
\label{Basis}
\end{eqnarray}
and their complex conjugates, where $k$ takes on suitable discrete values so
that periodic boundary conditions (in $x$) obtain.  Solutions of (\ref{Wave})
are thus well-behaved functions of $T$ even where $h=0$, at least in the sense
of one-sided limits.  Note that these are just the usual positive and negative
frequency solutions for $h<0$ and (anti)analytic functions of $x+i\sigma$ for
$h>0$ as expected.

\section{STOKES' THEOREM}

Now consider a manifold $M$ with a preferred hypersurface $\Sigma$ that
partitions $M$ into two manifolds-with-boundary $M^\pm$, and suppose that the
$n$-forms $\alpha^\pm$ are defined on $M^\pm$.  Let $V$ be an arbitrary region
of $M$, let $V^\pm$ denote the intersections $V \cap M^\pm$, and let
$(\partial V)^\pm = \partial V \cap (M^\pm{-}\Sigma)$.
Then we may use Stokes' Theorem (\ref{Stokes}) in $M^\pm$ to write
\begin{eqnarray}
\nonumber
\int_{V^+} d\alpha^+ + \int_{V^-} d\alpha^-
     &=&  \int_{(\partial V)^+} \alpha^+ + \int_{(\partial V)^-} \alpha^- \\
        && \quad + \int_\Sigma (\alpha^+ - \alpha^-)
\end{eqnarray}
where an assumption has been made about the relative orientation of $\Sigma$.
Consequently, if $\alpha^\pm$ are closed on $M^\pm$ and the pullbacks of
$\alpha^\pm$ to $\Sigma$ agree, then the standard arguments can be applied to
generate conserved quantities associated with the form $\alpha$ defined to be
$\alpha^\pm$ on $M^\pm$.%
\footnote{A related discussion of the divergence theorem in the presence of
signature change appears in \cite{Hellaby} which points out that conservation
of matter does not then follow from the junction conditions imposed on the
spacetime.}

We now assume that $\Sigma$ corresponds to the surface of signature change at
$t=0$ in the model of Section III, and that the wave equation (\ref{Wave}) is
satisfied on $M^\pm$.  In order for the (integral) momentum associated with
the Killing vector $\partial_x$ to be conserved, we need $J_\XX$ to satisfy
the additional condition above, namely that the pullbacks from $M^\pm$ to
$\Sigma$ agree.  But away from $t=0$
\begin{eqnarray}
\nonumber
J_\XX
  &=& \pm \left( {\Phi,_t{}^2 \over h} - \sgn(h) \Phi,_x{}^2 \right)
        \sqrt{|h|}\,dt \\
      && \quad \pm 2 {\Phi,_x \Phi,_t \over \sqrt{|h|}} dx
%  &=& \pm \left( \Phi,_T{}^2 - \sgn(h) \Phi,_x{}^2 \right) dT
%       + \pm \Phi,_x \Phi,_T dx \\
\end{eqnarray}
so that the pullback is $\pm 2 \Phi,_x \Phi,_T dx$, where the sign depends on
the choice of orientations of $M^\pm$, and the condition becomes
\begin{equation}
\Phi,_\sigma\AtSigma = \pm \Phi,_\tau\AtSigma
\label{TauMatch}
\end{equation}
where the sign now depends only on the relative orientation.
As shown in \cite{PaperII}, this requirement together with the natural
requirement that $\Phi$ be continuous at $\Sigma$ uniquely determine the
propagation of a solution of (\ref{Wave}) across $\Sigma$.

It is interesting to note that although (\ref{TauMatch}) implies that the
pullback condition is also satisfied for $K$, so that Klein-Gordon products
are automatically conserved, the converse is not true.

\section{DISTRIBUTIONS}

The massless wave equation for a 0-form $\Phi$ on a manifold with a
non-degenerate metric may be written
\begin{equation}
dF=0
\label{FWave}
\end{equation}
where
\begin{equation}
F=*d\Phi
\end{equation}
in terms of the Hodge map $*$ defined by the metric.  We wish to extend the
theory of the massless scalar field to manifolds that admit a metric that
changes signature. We sketch here how this may be achieved using a
distributional language and refer to \cite{PaperIV} for further  mathematical
details. We naturally require that (\ref{FWave}) be satisfied on
$\upm=\mpm{-}\s$, and we shall assume that $\Phi$ is continuous.

We shall call a 1-form $F$ on $M$ {\it regularly discontinuous} if the
restrictions $F^\pm=F|_\upms$ are smooth and the (1-sided) limits
$F^\pm\AtSigma=\lim_{t\to0^\pm} F$ exist.  The {\it discontinuity} of $F$ is
the tensor distribution on $\s$ defined by
\begin{equation}
[F]\OnSigma=\fp-\fm.
\end{equation}
Denote by $\ypm$ the Heaviside distributions with support in $\upm$ and such
that
\begin{equation}
d\ypm=\pm\d
\end{equation}
where $\d$ is the hypersurface Dirac distribution with support on $\s$.%
\footnote{The properties of these distributions will be discussed more fully in
\cite{PaperIV}. }
Now introduce, as a distribution on $M$,
\begin{equation}
\bff = F^+\bfy^+ + F^-\bfy^-
\end{equation}
It follows that
\begin{equation}
d\bff=\bfy^{+} dF^+ + \bfy^{-} dF^- + \d\wedge [F]\OnSigma.
\end{equation}
We readily deduce the consequences of requiring $d\bff$ to be the zero
distribution. By evaluating $d\bff$ on a set of test vectors with support
in $U^\pm$ we deduce
\begin{equation}
d F|_{\scriptscriptstyle U^\pm} =0
\end{equation}
as expected.
Similarly it follows that
\begin{equation}
\d\wedge[F]\OnSigma=0.
\label{DMatch}
\end{equation}

In order to derive the junction conditions for matching derivatives at $\s$ we
shall only admit solutions such that $F$ is regularly discontinuous at $\s$ so
that $[*d\Phi]\OnSigma$ is well defined. We seek distributional solutions to
\begin{equation}
d\bff=0
\end{equation}
where $\bff$ is defined as above with
$F|_\upms=*d\Phi|_\upms$.  Turning to the metric (\ref{Metric}), we
define
\begin{equation}
*1|_\upms=\epsilon^\pm \smh dt\wedge dx \, \Big|_\upms
\end{equation}
where $\epsilon^\pm=\pm1$ according to the choice of orientation of the
volume forms in $\upm$.  Then
\begin{equation}
*d\Phi_\pm=\pm \epsilon^\pm \left(
           {\p_t\Phi d x\over\smh}-\sgn(h)\p_x\Phi \smh dt \right) \Bigg|_\upm.
\end{equation}
If $\p_t\Phi/\smh$ is bounded as $t\to0^\pm$ then, since $\Phi$ is
assumed continuous, $*d\Phi$ is regularly discontinuous at $\s$ and
\begin{equation}
[*d\Phi]\OnSigma\wedge dt=\pm \left(\lim_{t\to 0^+}{\p_t\Phi\over\smh}-
              \epsilon \lim_{t\to 0^-}{\p_t\Phi\over\smh}\right)dx\wedge dt
\end{equation}
where $\epsilon=\epsilon^+/\epsilon^-$.  Then (\ref{DMatch}) implies
\begin{equation}
[*d\Phi]\OnSigma\wedge dt=0,
\end{equation}
which provides junction conditions for regularly discontinuous solutions of
the equations
\begin{equation}
d{*}d\Phi|_\upms=0.
\label{DWave}
\end{equation}
Furthermore, these junction conditions are identical to (\ref{TauMatch}).

\section{VARIATIONAL APPROACH}

It is not mandatory to use distributions to generate junction conditions. We
offer a variational approach that yields the same results for regularly
discontinuous forms. Consider the action
\begin{equation}
S[\Phi] = {1\over2}\int_\up d\pp\wedge *d\pp +
            {1\over2}\int_\um d\pmm\wedge *d\pmm
\label{Lagrangian}
\end{equation}
where the fields and metric are piecewise continuous and the appropriate Hodge
maps are understood in the regions $\upm$. Consider field variations
$\ppm\to\ppm+\ll\ppm$
\begin{eqnarray}
\nonumber
\delta S &=& \int_\up d(\ll\pp)\wedge *d\pp + \int_\um d(\ll\pmm)\wedge
*d\pmm\\
\nonumber
         &=& \int_{\p\up} \ll\pp{*}d\pp + \int_{\p\um}\ll\pmm{*}d\pmm \\
             && \quad + \int_\up\ll\pp d{*}d\pp +\int_\um\ll\pmm d{*}d\pmm
\end{eqnarray}
using Stokes' Theorem in $\upm$.  Now postulate that $\delta S=0$ for all
field variations of compact support.  Choosing support entirely in $\upm$
yields (\ref{DWave}).  Now let the variation have compact support on any
domain that includes $\Sigma$ and assume that $*d\Phi$ is regularly
discontinuous with respect to $\Sigma$.  Then the continuity of $\Phi$ allows
us to write $\delta\Phi=\delta\Phi^+\equiv\delta\Phi^-$ so that
\begin{equation}
\int_\Sigma \,\ll\Phi (*d\Phi^+- *d\Phi^-) = 0
\end{equation}
Since $\ll\Phi$ is arbitrary we conclude that the pullback of the form $
[*d\Phi]_\Sigma $ to the hypersurface $\Sigma$ must vanish.
If $\Sigma$ is given by $\psi=0$ then this may be expressed
in terms of a restriction
\begin{equation}
[*d\Phi]_\Sigma\wedge d\psi = 0
\end{equation}
as before.

The above argument can be generalized to other field theories.  We plan to
discuss the junction conditions for spinor fields in a discontinuous or
degenerate metric elsewhere; see also the recent work of Romano \cite{romano}.
We also observe that hypersurface sources are readily accommodated within this
language by considering actions of the form
\begin{equation}
S=\int_\up\Lambda_{+}+\int_\um\Lambda_{-}+\int_\Sigma\Lambda
\end{equation}
where the source is described by the hypersurface action density $\Lambda$.

\section{DISCUSSION}

It is important to distinguish between the assumptions made in our three
derivations of the matching condition (\ref{TauMatch}).  The first
derivation based on Stokes' theorem used conservation of momentum in the
presence of a Killing vector but did {\bf not} make use of a particular form
of the wave equation at the surface of signature change.  On the other hand,
the second derivation assumes a particular distributional form for the wave
equation on the whole manifold, while the third assumes a particular form for
the action; neither makes any assumptions about symmetries.  The latter two
derivations can therefore be applied in more general spacetimes provided one
is willing to accept either (\ref{FWave}) as being the correct wave
equation or (\ref{Lagrangian}) as being the correct action.  We plan to
apply this  approach to an explicit imbedding of the trousers spacetime in
3-dimensional Minkowski space.

If we assume that $f$ in (\ref{Metric}) has precisely 2 (simple) roots
corresponding to the $T$ values $T_i$ and $T_f$, and that $f<0$ as
$|t|\to\infty$, so that our spacetime is asymptotically Lorentzian, then, as
claimed in \cite{PaperI}\ and shown in detail in \cite{PaperII}, the above
solutions, satisfying the condition (\ref{TauMatch}) and continuity,
correspond to the relationship
\begin{eqnarray}
\nonumber
\Uout_k e^{+i|k|T_f} &=&
    \Uin_k e^{+i|k|T_i} \cosh(k\Delta T) \\
    && \quad + \bar \Uin_{-k} e^{-i|k|T_i} \> i \,\sinh(|k|\Delta T)
  .
\label{Results}
\end{eqnarray}
between basis solutions at early and at late times.  The mixing of positive
and negative frequencies, and hence particle production, is controlled by the
last term.

We note an interesting freedom in the derivations of the junction conditions
presented above.  In all three derivations, the choice of Hodge map in the
regions separated by $\s$ is fixed only up to a relative sign.  Physically,
this corresponds to different choices of time orientation in one or more
regions.  For the example just considered with two surfaces of signature
change, there are 8 different choices of orientations.  Since our (classical)
theory is invariant under a global change of time orientation, this number is
immediately reduced to 4.  (However, one might want on physical grounds to use
different boundary conditions depending on the global choice of time
orientation.)  Furthermore, it can easily be shown \cite{PaperII}\ that
changing the ``time'' orientation of the middle, Euclidean region results only
in an unimportant phase factor in (\ref{Results}), so that there is no need to
worry about which ``time'' orientation to pick in this region.  (Specifically,
the second term picks up a minus sign.)  Equation \ref{Results} corresponds
physically to a model with asymptotic ``in'' and ``out'' regions.  The only
remaining distinct choice corresponds to both Lorentzian regions being to the
future (or past) of the Euclidean region, corresponding to two universes
sharing a common Big Bang or Big Crunch.

A related case of interest is a paraboloid, e.g.\ with the induced metric
obtained from imbedding it in Minkowski 3-space with the rotation axis being
the time axis.  Deleting the point on the axis yields a manifold with topology
${\Bbb R}\times{\Bbb S}$ and a metric of the form (\ref{Metric}), but with
only one signature change, from an initial Euclidean region to a final
Minkowskian region.  This picture is reminiscent of quantum cosmology, and is
related to the models recently considered by Ellis {\it et al}.\ \cite{ellis}
and Hayward \cite{hayward}.  Note that there will be now be an extra
regularity condition at the axis which will affect the observed particle
spectrum at late times.

\nonumber
\section{ACKNOWLEDGMENTS}

It is a pleasure to thank George Ellis, Steve Harris, David Hartley, Sean
Hayward, Charles Hellaby, Joe Romano, Abe Taub, and Phillip Tuckey for helpful
conversations.  RWT thanks Oregon State University for hospitality while part
of this work was done. This work was partially funded by NSF grant
PHY 92-08494 (CAM \& TD).

\end{document}